\documentclass{llncs}
\usepackage{bbding}
\usepackage{amssymb}
\usepackage{graphicx}
\usepackage{epstopdf}
\usepackage{subfigure}
\usepackage{balance}
\usepackage{xcolor}
\usepackage{multirow}
\usepackage{cite}
\usepackage{algorithm}
\usepackage{algorithmic}
\usepackage{listing}
\usepackage{eqparbox}

\begin{document}
	\title{Modeling Conceptual Characteristics of Virtual Machines for CPU Utilization Prediction}
	\author{Shengwei Chen \and Yanyan Shen\thanks{corresponding author} \and Yanmin Zhu$^*$}
	\institute{Department of Computer Science and Engineering\\Shanghai Jiao Tong University\\
		\email{\{shineway\_chan, shenyy, yzhu\}@sjtu.edu.cn}}
	\maketitle
	\begin{abstract}
		Cloud services have grown rapidly in recent years, which provide high flexibility for cloud users to fulfill their computing requirements on demand. To wisely allocate computing resources in the cloud, it is inevitably important for cloud service providers to be aware of the potential utilization of various resources in the future. 
		This paper focuses on predicting CPU utilization of virtual machines (VMs) in the cloud.
		We conduct empirical analysis on Microsoft Azure's VM workloads and identify important conceptual characteristics of CPU utilization among VMs, including locality, periodicity and tendency. We propose a neural network method, named Time-aware Residual Networks (T-ResNet), to model the observed conceptual characteristics with expanded network depth for CPU utilization prediction. 
		We conduct extensive experiments to evaluate the effectiveness of our proposed method and the results show that T-ResNet consistently outperforms baseline approaches in various metrics including RMSE, MAE and MAPE.
	\end{abstract}

	\section{Introduction}\label{sec:intro}
	Recent years have witnessed the rapid growth of cloud computing technology. Many companies have migrated their workloads to cloud service platforms such as Microsoft Azure, Alibaba Cloud Compute Services, and Amazon Web Services. Under the pressure of market competition, cloud service suppliers have to provide attractive features to the customers while saving their platform costs, which, however, can be extremely hard to achieve without effective resource management.
	From the perspective of cloud resource management, understanding future demands of VM resources can help system administrators reallocate resources wisely in a dynamic manner. When it is foreseen that the demands for resources will increase, cloud providers could prepare more physical hosts to meet the growth of future demands in time. Similarly, when the demands of VM resources are predicted to experience a declining trend, cloud managers could stop allocating new resources and migrate the underloaded VMs properly so that the idle physical hosts can be shut down to avoid waste of resources and improve the lifetime of equipments. Among various resources for VM workloads, CPU utilization is one of the most important indicators since it has a great impact on the total cost of the cloud service~\cite{gusev2013cpu}. And it is more useful to understand the behaviors of maximum CPU utilization than average one, since the former keeps performance at high percentiles of the response time~\cite{herodotou2011starfish, clark2005live}. 
	
	In this paper, we focus on predicting maximum CPU utilization for long-running VMs based on historical utilization series. The key challenge of this problem lies in the instability of utilization series for each VM. Specifically, the series itself involves nonlinear short-term trends which increase the prediction difficulty. Furthermore, 
	the volatility of the series varies with time, and it is hard to capture long-term temporal dependencies without delicately designed prediction methods.
	Most existing works on CPU utilization prediction leverage the conventional machine learning methods such as ARIMA~\cite{calheiros2015workload}, linear regression~\cite{farahnakian2013lircup}, hidden Markov model (HMM)~\cite{gong2010press, khan2012workload}, and multilayer perceptron model (MLP)~\cite{islam2012empirical}. However, the performances of these methods are far from satisfactory. In particular, linear regression and ARIMA can only capture linear relationships for time series due to the restricted model complexity, and HMM-based methods can only predict the change of patterns according to pre-defined finite states. MLP can forecast nonlinear relationships but the model itself is too simple to catch long-term temporal dependencies.
	
	To provide an effective method for CPU utilization prediction, we first conduct empirical analysis on real CPU utilization dataset from Microsoft Azure~\cite{cortez2017resource}, from which we obtain two important observations and extract key concepts as follows. 
	First, CPU utilizations for VMs within one deployment often fluctuate together over time due to the fact that multiple VMs in the same deployment are typically created to execute tasks collaboratively.
	Second, VM CPU utilization as a time series presents three conceptual characteristics: (1) locality: VM CPU utilization at present will impact the value in the near future; (2) periodicity:  utilization series shows cyclical changes over time; (3) tendency: utilization will continue to increase or decrease in the long run with the change of work intensity. 
	These refined characteristics as part of CPU utilization reflect the user's real-time requirements from different aspects.
	
	Based on the above key observations, we develop a neural network method to solve the maximum VM CPU utilization prediction problem in consideration of VM behaviors. Given a target VM, we first use the Pearson correlation coefficient to select the most relevant VMs' utilization from a deployment, which are used as additional inputs to expand the target utilization series. We then divide the expanded utilization series into three parts at different time frequencies to represent locality, periodicity, tendency properties, respectively. Finally, we propose Time-aware Residual Networks (T-ResNet) based on residual networks~\cite{he2016deep}. Our T-ResNet takes three divided utilization series as inputs, and models each component by using an individual residual networks. The residual structure is able to capture both nonlinear short-term volatility and the long-term temporal dependencies simultaneously. 
	The outputs of sub-networks are concatenated and fed to a fully-connected layer to re-weight the importance of latent features from different residual networks for final maximum CPU utilization prediction.
	The experimental results on Microsoft Azure data show that our model outperforms five benchmark methods in various metrics.
	
	The contributions of this paper can be summarized as follows:
	\begin{itemize}
		\item We analyze the real Microsoft Azure dataset to disclose important conceptual characteristics of VM CPU utilization. We exploit both stochastic behaviors of VM CPU utilization as well as CPU utilization similarity among VMs.
		\item We propose Time-aware Residual Networks (T-ResNet) for CPU utilization prediction. T-ResNet dynamically aggregates the outputs of residual networks which model locality, periodicity, and tendency respectively, assigning different weights to different frequency patterns. The aggregation is further activated to generate utilization prediction.
		\item We conduct extensive experiments to evaluate the performance of our proposed method. The results show that our model outperforms five baseline approaches by reducing 19\%-64\% in RMSE, 38\%-106\% in MAE, 43\%-132\% in MAPE.
	\end{itemize}
	
	{\bf Organization.} Section~\ref{sec:preliminaries} provides definitions and problem statement. Section~\ref{sec:empirical} presents the data set and empirical analysis for VM CPU utilization. Section~\ref{sec:t-resnet} introduces the proposed model. Section~\ref{sec:experiment} gives experimental results. We review related works in Section~\ref{sec: related} and conclude this paper in Section~\ref{sec:conclusion}.
	
	\section{Preliminaries} \label{sec:preliminaries}
	\begin{definition}[VM CPU Utilization]\label{def:vm-utilization}
		The virtual machine is the smallest executable unit hosted on physical servers. 
		Let $\mathcal{V}=\{v_t | t=1,\cdots,T\}$ be a series of CPU utilization for one VM, where $v_t$ is a 3-dimensional vector containing minimum, average, and maximum CPU utilization of the VM at time period $t$, denoted by $v_{min,t}$, $v_{avg,t}$ and $v_{max,t}$, where the time interval is fixed, e.g., 5 minutes.
	\end{definition}
	\begin{definition}[Deployment]\label{def:deployment}
		A deployment contains a set of VMs that are managed together and allocated in the same cluster of servers. Let $\mathcal{D}=\{\mathcal{V}^{i} | i=1,2,\cdots,N\}$ be a deployment with $N$ VMs, where the CPU utilization of the $i$-th VM is represented by $\mathcal{V}^i$. 
		Note that the VMs in the same deployment typically collaborate with each other for executing the same task.
	\end{definition}

	Since users prefer to execute tasks in batches by increasing (or shrinking) the number of VMs in a deployment, we focus on the deployments which contain VMs with long-running workloads in this paper. The prediction of CPU utilization of any newly created VMs is beyond the scope of this paper. Then we give the problem definition we would like to solve. 
	
	\noindent {\bf Problem statement.}
	Given a deployment $\mathcal{D}=\{v_{t}^{i} | i\in[1,N]\wedge t\in[1,T]\}$ where $v_{t}^{i}$ denotes the CPU utilization of the $i$-th VM during the $t$-th time interval, we aim to predict the maximum CPU utilization of all the VMs in $\mathcal{D}$ at time period $T+1$, i.e., $v_{max, T+1}^{i}$ for all $ i\in[1,N]$.
	
	\section{Empirical Analysis} \label{sec:empirical}
	We start with the description of the Microsoft Azure dataset\footnote{\url{https://github.com/Azure/AzurePublicDataset}} offered by Cortez et al.~\cite{cortez2017resource}. 
	We then perform an empirical analysis on the dataset and present several key observations on CPU utilization behaviors of the VMs within a deployment.
	
	\subsection{Microsoft Azure Dataset} \label{ssec: Dataset}
	{\bf Data description.}
	Microsoft Azure is one of the largest and most influential cloud providers in the world, and contains both first-party and third-party VM workloads. The dataset from~\cite{cortez2017resource} only includes first-party workloads of more than two million VMs running on Azure spanning cross 30 consecutive days from November 16 to December 15, 2016. The first-party workloads mainly combine Microsoft development, test, and internal services.
	
	\noindent {\bf Data preprocessing.}
	As we aim at predicting CPU utilization of long-running VMs, we first filter VMs with short lifetime from the original dataset. In this paper, we consider deployments where all VMs run throughout the entire life cycle of the dataset, i.e., 30 consecutive days. After filtering deployments that do not satisfy our requirement, we obtain 3,005 deployments with 16,065 VMs. And the following presented observations are based on these remaining deployments.

\subsection{Stochastic Behaviors of VM CPU Utilization} \label{ssec: Intra-VM}
Since VM may execute tasks for a long time, it is meaningful to analyze the CPU utilization patterns belonging to each VM itself. 
Seasonal decomposition is a statistical method used for time series decomposition, which decomposes a time series into several components: trend, seasonality, and residual~\cite{cleveland1990stl}. The trend is defined as the increasing or decreasing value in the series; seasonality is defined as the repeating short-term period in the series; residual is defined as the random variation in the series. Based on seasonal decomposition, we are able to transform the CPU utilization series into several components as follows:
\begin{equation}
v_{max,t} = vt_{t} + vs_{t} + vr_{t}
\end{equation}
where $v_{max,t}$ is the maximum CPU utilization during the $t$-th time interval, and $vt_{t}$, $vs_{s}$, $vr_{t}$ are the respective decomposed components. 
Seasonal decomposition adopts smoothing technique to calculate the trend.
Next, we estimate seasonality by averaging the de-trended values for a specific season. Finally, we get residual component by removing the trend and seasonality from the original series.

\begin{figure}[t]
	\centering
	\includegraphics[width=.9\linewidth]{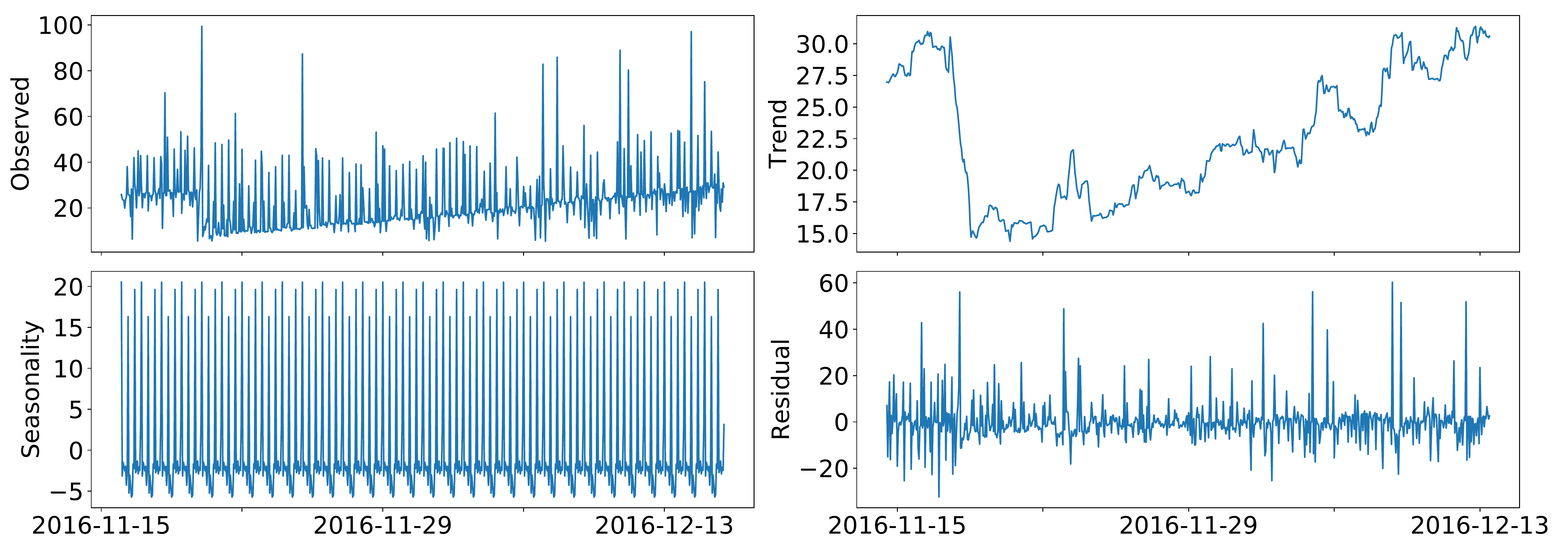}
	\vspace{-.1in}
	\caption{Decomposed maximum CPU utilization by seasonal decomposition}
	\label{fig:Intra-Seasonal}
	\vspace{-.1in}
\end{figure}

Fig.~\ref{fig:Intra-Seasonal} illustrates the original CPU utilization series and its decompositions for a sampled VM. From the figure, we observe the following conceptual characteristics in terms of behaviors inside VM:
\begin{itemize}
	\item \textit{Locality}: CPU utilization is continuously changing over time and hence we should consider continuous utilization data in 5-minutes granularity, which reflects the short-term dependencies that tend to be similar.	
	\item \textit{Periodicity}: The de-trended utilization series shows stable periodicity every day, and this reveals some workloads consume CPU resource in a diurnal cycle.
	\item \textit{Tendency}: There exists an increasing trend in CPU utilization after a sudden drop.
\end{itemize}
Inspired by the above observations, we explicitly derive three key fragments from the original 5-minutes CPU utilization series, to capture \textit{locality}, \textit{periodicity} and \textit{tendency} behaviors, respectively. This is achieved by sampling data at 5-minutes, one-hour and one-day granularities.

\subsection{CPU Utilization Similarity among VMs} \label{ssec: Inter-VM}
\begin{figure}[t]
	\centering
	\includegraphics[width=.9\linewidth]{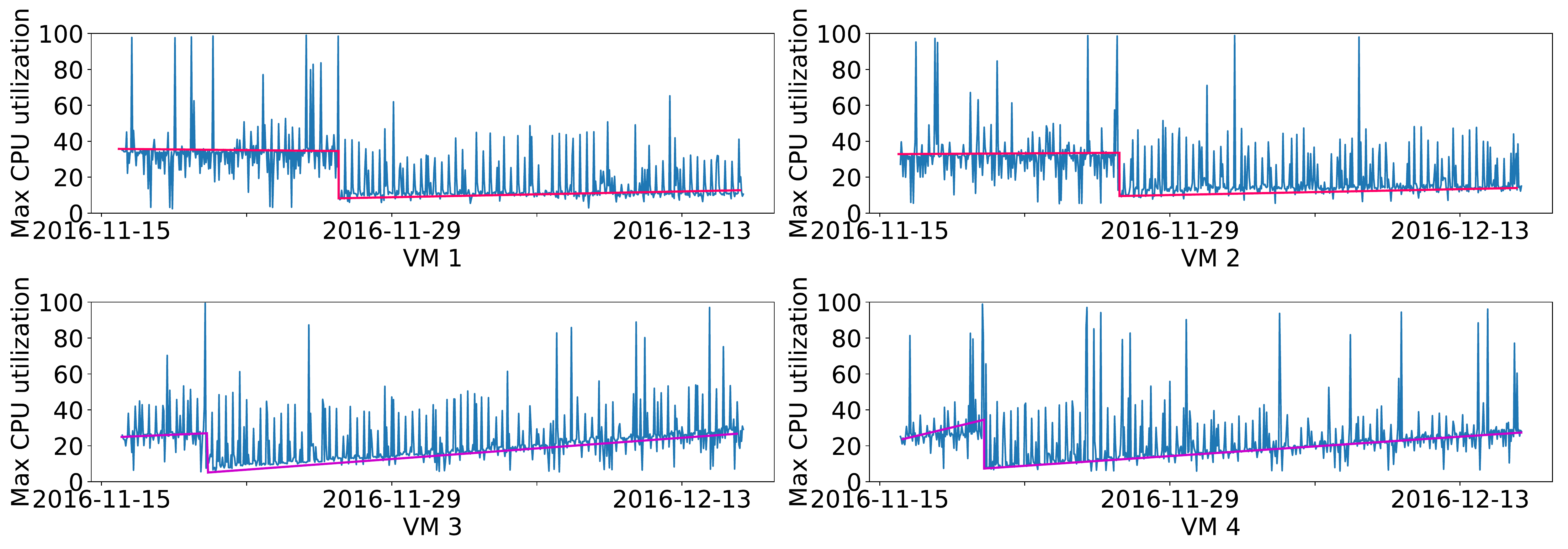}
	\vspace{-.1in}
	\caption{Maximum CPU utilization on four VMs in the same deployment}
	\label{fig:Inter-VM}
	\vspace{-.1in}
\end{figure}

As we mentioned above, some VMs deployed in the same deployment tend to execute the same type of tasks and exhibit some common patterns. 
To better visualize the relationships between different VMs in the same deployment, we downsample the origin time series from five minutes granularity into one hour granularity by selecting a maximum point at one-hour interval. 
Fig.~\ref{fig:Inter-VM} shows the sample maximum CPU utilization time series of 4 VMs collected in the same deployment covering a consecutive month.
From Fig.~\ref{fig:Inter-VM}, we can see that: VM1 and VM2 appear to fluctuate around an average line (marked in pink) and there is a trigger that makes the average CPU utilization smaller; VM3 and VM4 fluctuate steadily before an abrupt decline, followed by a continuous upward trend (marked in purple).
These facts suggest that \textit{some VMs in the same deployment should be relevant, i.e., they work in a parallel and collaborative manner.}

\begin{figure}[t]
	\centering
	\includegraphics[width=0.65\linewidth]{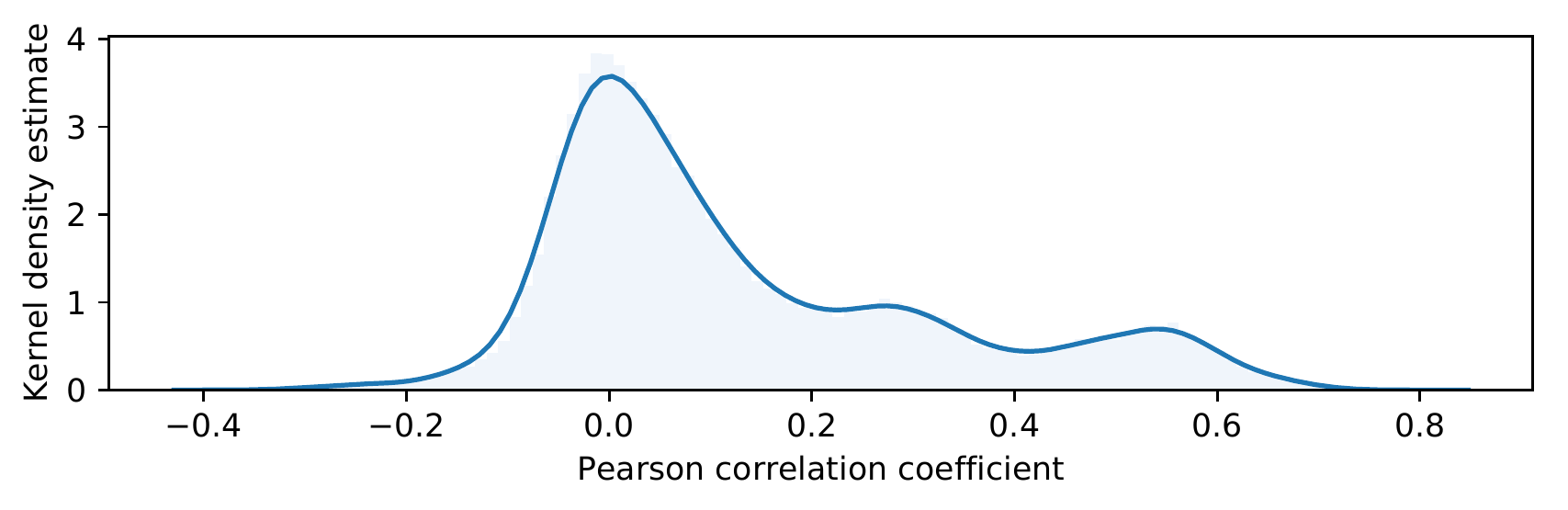}
	\vspace{-.1in}
	\caption{Kernel density estimate of Pearson correlation}
	\label{fig:KDE-Pearson}
		\vspace{-.1in}
\end{figure}

To measure the correlation between different VMs, we first randomly sample 320 VM CPU utilization series from the filtered deployments, and then calculate the Pearson correlation coefficient between different VMs. 
Fig.~\ref{fig:KDE-Pearson} shows the Gaussian Kernel Density Estimate (KDE)~\cite{silverman2018density} plot of pairwise VM correlation with setting bins as 100. 
The $x-$axis is the Pearson correlation, and the $y-$axis is the Gaussian weighted sum of nearby densities. We can see that there exists a positive correlation for pairwise VMs.
To leverage this observation for improving prediction accuracy, we pick the maximum CPU utilization series of the $K$ VMs that are most relevant to the target VM as external inputs. Note that these CPU utilization series are aligned over the timeline.
Formally, we define the new CPU utilization series of target VM, as follows:
\begin{definition} [Expanded VM CPU utilization] \label{def: vm-expanded-utilization}
	Let $\mathcal{T}=\{x_{t} | t=1,2,\cdots,T\}$ be a series of expanded CPU utilization for one VM, where $x_{t}$ is a (3+$K$)-dimensional vector containing minimum, average and maximum CPU utilization of the target VM and maximum CPU utilization of other $K$ most relevant VMs.
\end{definition}

In what follows, we introduce our neural method that incorporates all the above observations for predicting CPU utilization in next time period.

\section{Time-aware Residual Networks for CPU Utilization Prediction} \label{sec:t-resnet}

\begin{figure}[t]
	\centering
	\includegraphics[width=0.8\linewidth]{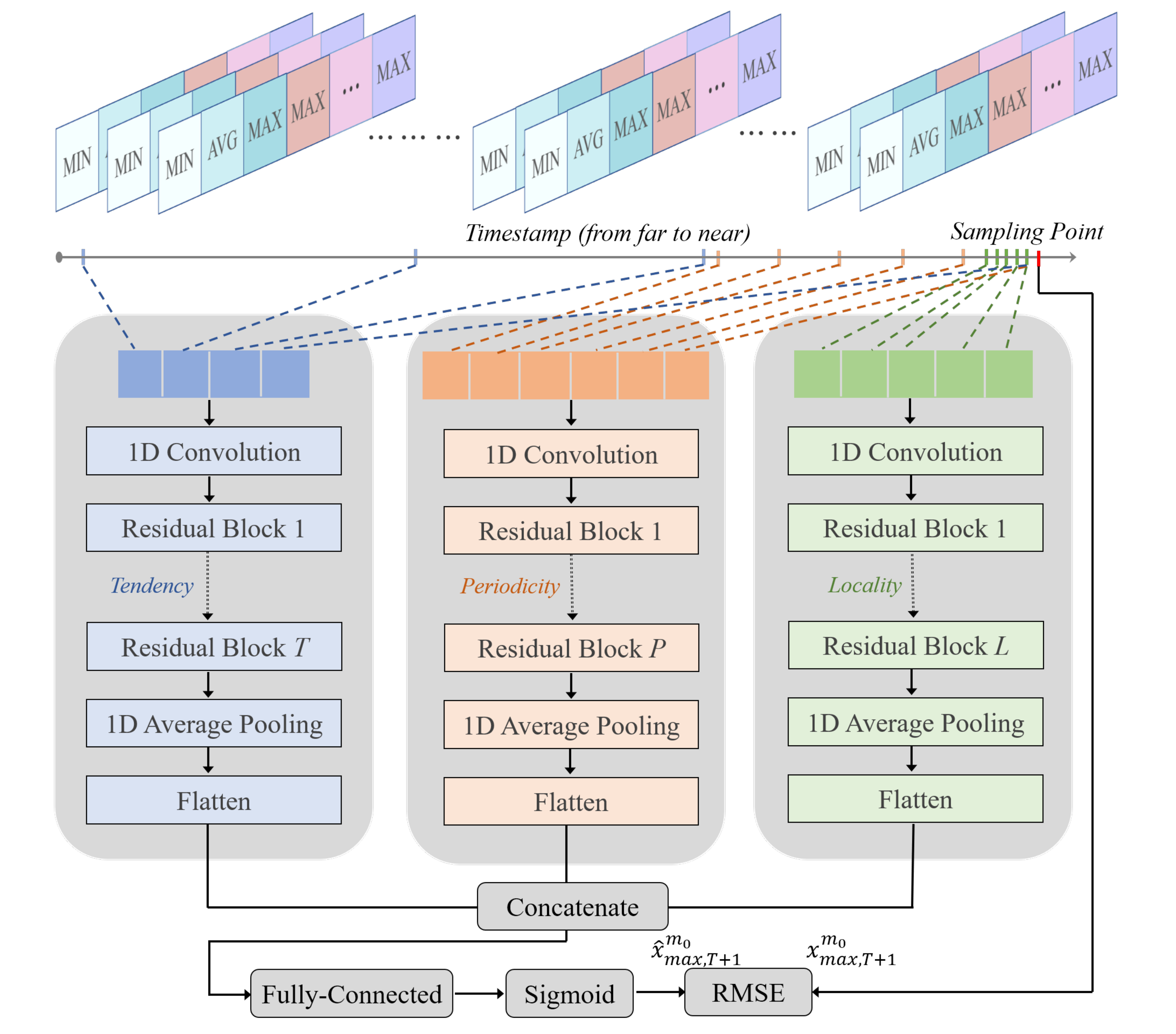}
	\vspace{-.1in}
	\caption{Time-aware Residual Networks}
	\label{fig:Time-aware}
	\vspace{-.2in}
\end{figure}
Inspired by the model in~\cite{zhang2017deep}, we propose our Time-aware Residual Networks (T-ResNet) model. Fig.~\ref{fig:Time-aware} depicts the structure of our T-ResNet, which contains three major components modeling temporal \textit{locality}, \textit{periodicity}, and \textit{tendency}, respectively. As shown in top part of Fig.~\ref{fig:Time-aware}, we first choose $K$ VMs that are most relevant to the target VM based on Pearson correlation coefficient and expand the target VM CPU utilization according to Definition~\ref{def: vm-expanded-utilization}. We then split the expanded utilization series from time axis into three fragments by sampling at 5-minutes, one-hour, and one-day frequencies. After that, the sampled 2D tensors of each fragment are fed into three residual networks accordingly to model \textit{locality}, \textit{periodicity}, and \textit{tendency}, respectively. Finally, the flattened outputs of three residual networks are concatenated and further put into the fully-connected neural network to generate the predicted maximum CPU utilization value for the target VM in the next time period. Since the CPU utilization ranges from 0 to 1 after Min-Max normalization, we use the sigmoid function to activate final output and try to minimize the loss between true and predicted values through backward propagation~\cite{rumelhart1985learning} during model training.

\begin{figure}[t]
	\centering
	\includegraphics[width=0.55\linewidth]{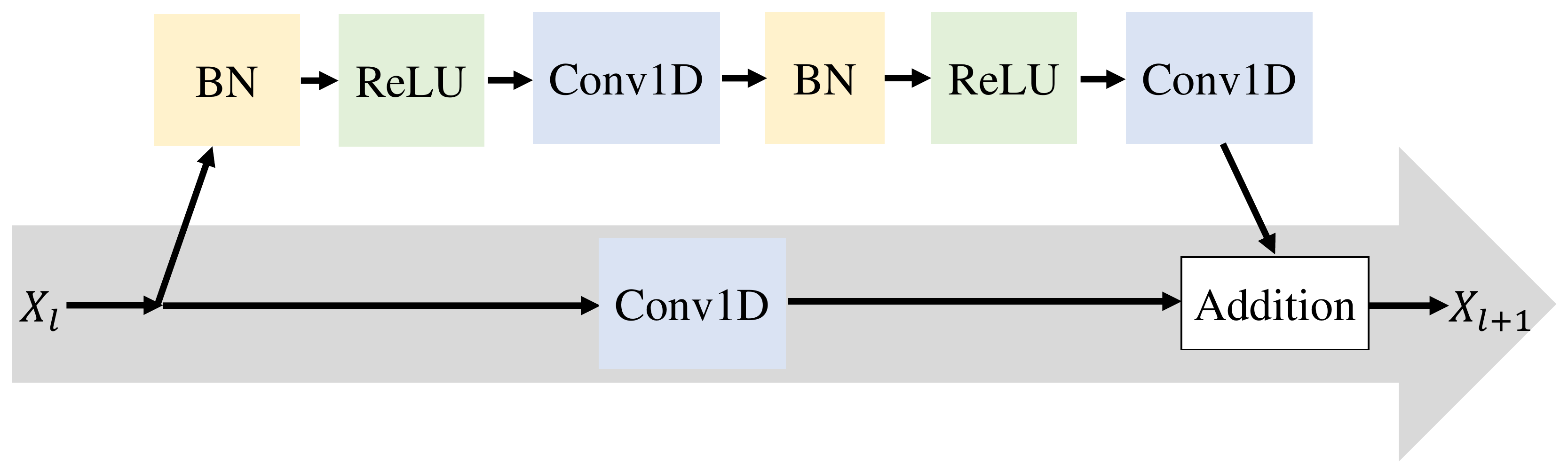}
	\vspace{-.1in}
	\caption{Residual block. BN: Batch Normalization; Conv1D: 1-D Convolution}
	\label{fig:ResBlock}
	\vspace{-.15in}
\end{figure}

The structure of each residual network in Fig.~\ref{fig:Time-aware} is composed of convolution and the residual block.
Convolutional neural network (CNN) is first proposed to handle image recognition problem by Yann et al.~\cite{lecun1998gradient}, and it can also be used for applications other than image recognition such as signal processing~\cite{van2016wavenet} and time series classification~\cite{cui2016multi}. From Section~\ref{ssec: Intra-VM}, we observe repeatable local patterns in utilization series which could be detected via the convolutional operation. 
As single convolution captures local patterns, stacking multiple convolutions together can identify much more complex patterns and capture global temporal dependencies of VM utilization. However, deeper networks were difficult to train due to the notorious problem of gradient vanishing/exploding which blocks convergence~\cite{glorot2010understanding}. Fortunately, the residual network solves this problem by adding shortcut connection to residual block (stacking of two layers of CNN, as shown in Fig.~\ref{fig:ResBlock}). Formally, the connection is defined as: $\mathbf{y}=\mathbf{W}\mathbf{x}+\mathcal{F}\left( \mathbf{x} \right)$,
where $\mathbf{x}$, $\mathbf{y}$ are inputs and outputs, respectively. $\mathbf{W}$ represents linear transformation and $\mathcal{F}$ is residual function~\cite{he2016deep}. The key idea of the residual network is to learn residual function $\mathcal{F}$. Therefore, we employ residual networks with deep structures to capture global temporal dependencies.

We develop T-ResNet to predict CPU utilization for all the VMs in a deployment. For illustration purpose, we describe the procedure of predicting one VM CPU utilization called target series. 
The prediction of other VMs in the same deployment is performed in the same way.
Assume we select $K$ most relevant VMs, and the extended target CPU utilization series can be expressed as $\mathcal{T}=\{x_{t} | t=1,2,\cdots,T\}$, according to Definition~\ref{def: vm-expanded-utilization}. For the residual network of \textit{locality} in Fig.~\ref{fig:Time-aware}, we have the input fragment to be $\left[x_{T-l_{l}+1}, x_{T-l_{l}+2}, \cdots, x_{T}\right]$, where $l_{l}$ is the length of intervals of \textit{locality} component to look back. Then we concatenate them along time axis to be a two-dimensional tensor $\mathbf{X}_{0}^{l} \in \mathbf{R}^{l_l \times (3+K)}$. We use 1-D convolution to extract shallow characters (see Fig.~\ref{fig:Time-aware}):
\begin{equation}
\small
	\mathbf{X}_{1}^{l}=W_{1}^{l}*\mathbf{X}_{0}^{l}
\end{equation}
where $*$ denotes the convolution operation, $W_{1}^{l}$ denotes the filters of the first layer, and biases are omitted for simplifying notations. After that, we use multiple stacking residual blocks to model global temporal relationship. Fig.~\ref{fig:ResBlock} shows the residual blocks used in this paper, which can handle the dimension inequality problem in shortcut~\cite{he2016identity}. Each residual block (i.e., the upper branch in Fig.~\ref{fig:ResBlock}) involves two sets of ``Batch Normalization + ReLU + Convolution''. Batch Normalization accelerates deep network training by reducing internal covariate shift~\cite{ioffe2015batch} and ReLU is a nonlinear activation function which helps network capture more complex patterns~\cite{nair2010rectified}. In order to extract features as much as possible and reduce information loss, we convolve to halve the dimension of time step while doubling the dimension of features by adding filters. To add the origin inputs of the residual network to transformed outputs, we perform a linear projection on the shortcut connection to match the dimension. In short, the layer $k$ of the residual block can be formulated as:
\begin{equation}
\small
	\mathbf{X}_{k}^{l} = \mathcal{F}\left(\mathbf{X}_{k-1}^{l},\{W_{k,1}^{l},W_{k,2}^{l}\}\right)+W_{k,3}^{l}*\mathbf{X}_{k-1}^{l}
\end{equation}
where $\mathcal{F}$ represents the residual mapping, i.e., $	\mathcal{F}=W_{k,2}^{l}*\sigma\left(W_{k,1}^{l}*\sigma\left(\mathbf{X}_{k-1}^{l}\right)\right)
$; $\sigma$ denotes ReLU; $W_{k,1}^{l}$ and $W_{k,2}^{l}$ denote filters; $W_{k,3}^{l}$ are filters used for dimension transform. We omit batch normalization for simplicity. Noticeably, these time steps of inputs of our three residual networks may be different and will shrink as the depth deepen, hence the depth of networks required to reduce the dimension of time steps varies. For the residual network of \textit{locality}, we stack $L$ residual blocks upon the first layer to reduce the length of time steps to 2, and then get outputs $\mathbf{X}_{L+1}^{l}$. In addition, we use average pooling which can detect more background information and reduce noise followed by flattening to further transform features into one-dimensional vector denoted by $\vec{x}^{l}$.

Similarly, based on above operation, we can construct the residual networks of \textit{periodicity} and \textit{tendency} in Fig.~\ref{fig:Time-aware}. We represent the \textit{periodicity} fragment as $\left[x_{T-\left(l_{p}-1\right)*T_{p}}, x_{T-\left(l_{p}-2\right)*T_{p}}, \cdots, x_{T}\right]$, where $l_{p}$ denotes the length of intervals of period, and $T_{p}$ is the period span. After stacking $P$ residual blocks, the output of \textit{periodicity} is $\vec{x}^{p}$. Meanwhile, the output of \textit{tendency} is $\vec{x}^{t}$, given the inputs $\left[x_{T-\left(l_{t}-1\right)*T_{t}}, x_{T-\left(l_{t}-2\right)*T_{t}}, \cdots, x_{T}\right]$. $l_{t}$ is the length of \textit{tendency} fragment and $T_{t}$ denotes trend frequency, by stacking $T$ residual blocks. 

Finally, we concatenate $\vec{x}^{l}$, $\vec{x}^{p}$ and $\vec{x}^{t}$ together as a new vector $\vec{x}_{res}$, and feed it into a fully-connected layer to dynamically adjust weights of different features extracted from residual networks. Formally, we have:
\begin{equation}
\small
	\widehat{x}_{max, T+1}^{m_0}=\delta\left(W_{res}\vec{x}_{res}\right)
\end{equation}
where $m_0$ denotes target VM, $W_{res}$ is weights of the fully-connected neural network, and $\delta$ denotes the sigmoid activation function. We train our model by minimizing the mean square error (MSE), between true and predicted maximum utilization values, as described below:
\begin{equation}\label{eq:mse}
\small
	L = \frac{1}{N}\sum_{t=1}^{N}\left(\widehat{x}_{max,t}^{m_0}-x_{max,t}^{m_0}\right)^{2}
\end{equation}
where $N$ is the number of training samples.

\begin{algorithm}[t]
	\caption{Time-aware Residual Networks Training Algorithm}   
	\label{alg:training}   
	\small
	\begin{algorithmic}[1] 
		\REQUIRE ~~\\ 
		VM CPU utilizations in a deployment: $\{\mathcal{V}^1,\cdots,\mathcal{V}^{N}\}$, where $\mathcal{V}^{i}=\{v_1^i,\cdots,v_T^i\}$;\\
		Lengths of \textit{Locality}, \textit{Periodicity}, \textit{Tendency} fragments: $l_{l}$, $l_{p}$, $l_{t}$ ;\\ 
		Periodicity span: $T_{p}$; Tendency frequency: $T_{t}$; Number of relevant VM: $K$;\\
		\ENSURE ~~\\ 
		Learned Time-aware Residual Networks model \\
		\COMMENT{\textit{//generate new CPU utilization series}} \\
		\STATE Select $K$ most relevant VMs in the deployment and expand the origin series for each VM to get: $\{\mathcal{T}^{1},\cdots,\mathcal{T}^{N}\}$, where $\mathcal{T}^{i}=\{x_1^i,\cdots,x_T^i\}$ \\
		\STATE $\mathcal{U}\leftarrow\emptyset$ \\
		\FOR {\textit{all expanded VM CPU utilization $\mathcal{T}^{i}\left(i\in[1,N]\right)$}}
		\FOR {\textit{all available training timestamps $ts\left(ts\in[1,T]\right)$}}
		\STATE $\mathcal{F}_{l}=\left[x_{ts-l_{l}+1}^{i}, x_{ts-l_{l}+2}^{i}, \cdots, x_{ts}^{i}\right]$ \\
		\STATE $\mathcal{F}_{p}=\left[x_{ts-\left(l_{p}-1\right)*T_{p}}^{i}, x_{ts-\left(l_{p}-2\right)*T_{p}}^{i}, \cdots, x_{ts}^{i}\right]$ \\
		\STATE $\mathcal{F}_{t}=\left[x_{ts-\left(l_{t}-1\right)*T_{t}}^{i}, x_{ts-\left(l_{t}-2\right)*T_{t}}^{i}, \cdots, x_{ts}^{i}\right]$ \\
		\STATE $x_{max,ts+1}^{i}$ is the true maximum CPU utilization at next duration $ts+1$ \\
		\STATE add an training sample $\left(\{\mathcal{F}_{l},\mathcal{F}_{p},\mathcal{F}_{t}\},x_{max,ts+1}^{i}\right)$ into $\mathcal{U}$
		\ENDFOR
		\ENDFOR \\
		\STATE Initialize all the network parameters $\theta$ in T-ResNet, 
		\WHILE{$\theta$ \textit{not converged}}
		\STATE Select a batch of samples $\mathcal{U}_b$ randomly from $\mathcal{U}$ 
		\STATE Find parameters $\theta$ by minimizing the loss defined in Eq.~\ref{eq:mse} with $\mathcal{U}_b$
		\ENDWHILE
	\end{algorithmic}  
\end{algorithm}

Algorithm~\ref{alg:training} describes the preprocessing and training process of our T-ResNet method. We first pick $K$ most relevant VMs for each target VM to expand the origin series (line 1). We then construct the training samples via the whole deployment (lines 2-11). Finally, we train the T-ResNet based on Adam optimizer~\cite{kingma2014adam}, a variant of Stochastic Gradient Descent(SGD) (lines 12-16). Due to the fact that objective function is non-convex, the gradient-based optimization methods are usually trapped into the local optimum. Fortunately, Adam fuses the advantages of Momentum method ~\cite{polyak1964some} and RMSprop method~\cite{Tieleman2012} to overcome this problem. Specifically, Momentum considers the direction of last gradients, while RMSprop adopts the exponential moving average method to filter historical gradients. Such a combination effectively speeds up network learning process and helps the training process escape from local optimum.

\section{Experiments} \label{sec:experiment}
\vspace{-.06in}
\subsection{Experimental Setup}
\noindent \textbf{Datasets.}
We use the Microsoft Azure dataset for performance evaluation. As discussed in Section~\ref{ssec: Dataset}, we filter many deployments that do not sustain the entire lifetime of our dataset and leave 3005 deployments. After considering the rationality of experiments and the limitations of resources, we prepare two training sets, \textbf{Dep1} and \textbf{Dep2}, each of which includes 32 VMs in a deployment.

\noindent \textbf{Baselines.}
To demonstrate the effectiveness of T-ResNet, we compare it against 5 baseline methods.
\begin{itemize}
	\item \textbf{NA\"IVE}: We predict the maximum VM CPU utilization by the maximum CPU utilization of that VM at last time interval.
	\item \textbf{ARIMA}~\cite{makridakis1997arma}: Autoregressive Integrated Moving Average (ARIMA) model is a generalization of an autoregressive moving average (ARMA) model, which is widely used in time series analysis.
	\item \textbf{SARIMA}: Seasonal ARIMA which incorporates both non-seasonal and seasonal factors in ARIMA.
	\item \textbf{XGBoost}~\cite{chen2016xgboost}: XGBoost is a scalable machine learning system for tree boosting.
	\item \textbf{LSTM}~\cite{hochreiter1997long}: Long Short-Term Memory (LSTM) network are a special kind of recurrent neural network (RNN), which can learn long-term dependencies.
\end{itemize}

\noindent \textbf{Experimental Settings.}
We train each model using the whole training set, i.e., one deployment. Since the value of maximum CPU utilization varies greatly, we first transform origin series into log-scale to balance the order of magnitude. We then use Min-Max normalization to further scale data into the range $[0,1]$.
For all residual blocks after the first convolutional layer, we halve the length of time step by setting filter stride as 2 while doubling the feature dimension by doubling the filter numbers. 
All the filter size in convolution is set to 3. 
In our T-ResNet, $T_{p}$ and $T_{t}$ are empirically fixed to one-hour and one-day. Based on our observations, we set $l_{l}$, $l_{p}$, and $l_{t}$ to 12, 24, and 7, respectively.
Once the hyperparameter $K$ is determined, all the VMs in a deployment will select the same number of relevant VMs as additional inputs and we set $K\in\left\{0,2,4\right\}$. For each deployment, we select the first two-weeks data as the training set, the following week's data as the validation set, and the last week's data as the test set. The validation set is used for early stopping and selection of best network parameters.

\noindent \textbf{Metrics.}
To measure the effectiveness of various methods, we introduce three different evaluation measures. Among them, root mean square error (RMSE) and mean absolute error (MAE) are scale-dependent metrics, while mean absolute percentage error (MAPE) is scale-independent metrics. Specifically, the RMSE is defined as $\mathbf{RMSE}=\sqrt{\frac{1}{N}\sum_{t=1}^{N}\left(\hat{y}_t-y_t\right)^2}$, MAE is defined as $\mathbf{MAE}=\frac{1}{N}\sum_{t=1}^{N}|\hat{y}_t-y_t|$, and MAPE is defined as $\mathbf{MAPE}=\frac{1}{N}\sum_{t=1}^{N}|\frac{\hat{y}_t-y_t}{y_t}|$
, where $\hat{y}$ is the predicted value and $y$ is the true value.

\vspace{-.1in}
\subsection{Main Results}
\begin{table}[t]
	\caption{Comparison results over the Dep1 and Dep2 dataset (best performance displayed in \textbf{blodface}). Our original model does not consider relevant VMs, and $k$REL means adding extra most relevant $k$ VMs into the model.}
	\label{tab:Results}
	\centering
	\begin{tabular} {|c|c|c|c|c|c|c|}
		\hline \hline
		\multirow{4}{*}{Models} & \multicolumn{3}{c|}{\multirow{2}{*}{\textbf{Dep1 Dataset}}} & \multicolumn{3}{c|}{\multirow{2}{*}{\textbf{Dep2 Dataset}}} \\ 
		&\multicolumn{3}{c|}{ }&\multicolumn{3}{c|}{ } \\ \cline{2-7}
		& \textbf{RMSE} & \textbf{MAE} & \textbf{MAPE} & \textbf{RMSE} & \textbf{MAE} & \textbf{MAPE} \\ 
		&$\left(\times 10^{-2}\right)$&$\left(\times 10^{-2}\right)$&$\left(\times 10^{-2}\right)$&$\left(\times 10^{-2}\right)$&$\left(\times 10^{-2}\right)$&$\left(\times 10^{-2}\right)$ \\
		
		\hline \hline
		NA\"IVE & 11.87 & 6.95 & 16.75 & 11.96 & 7.03 & 18.10 \\ \hline
		ARIMA & 9.43 & 6.89 & 19.97 & 10.34 & 8.81 & 25.99 \\ \hline
		SARIMA & 10.16 & 6.63 & 16.08 & 10.77 & 7.05 & 18.06  \\ \hline
		XGBoost & 8.99 & 7.21 & 21.07 & 9.50 & 7.70 & 24.83 \\ \hline
		LSTM & 8.78 & 5.68 & 14.20 & 8.83 & 6.43 & 17.60 \\ \hline \hline
		T-ResNet & 7.68 & 4.66 & 11.37 & 7.81 & 4.78 & 12.32 \\ \hline
		T-ResNet-2REL & \textbf{7.35} & \textbf{4.11} & \textbf{9.95} & \textbf{7.28} & \textbf{4.27} & \textbf{11.19} \\ \hline
		T-ResNet-4REL & 7.62 & 4.71 & 12.41 & 7.68 & 4.47 & 11.76 \\ \hline \hline	
	\end{tabular}
	\vspace{-.06in}
\end{table}
The maximum CPU utilization prediction results of T-ResNet and other baseline methods over the two selected deployments are shown in Table~\ref{tab:Results}.

\noindent \textbf{Results of Compared Methods}
In Table~\ref{tab:Results}, we observe that the RMSE of NA\"IVE method is generally worse than other methods. This result shows that not all VMs tend to be stable in short term.
ARIMA and SARIMA are all linear regression with differential operation in nature~\cite{makridakis1997arma}. Both of them have slightly better performance compared with NA\"IVE method, however, the ability of them is limited which fails to capture nonlinear relationship. XGBoost is a tree-based model which combines classification and regression tree (CART) algorithm with gradient boosting method~\cite{chen2016xgboost}, hence it can capture more complicated relationship and perform better than ARIMA in RMSE. As shown, XGBoost achieves little improvement since it can only consider a few steps of inputs. LSTM is particularly designed to remember information for long periods of time due to the existence of memory cell, but it does not show significant performance improvement according to the results.
The reason may be that the above-mentioned key drivers of our dataset may not be captured by LSTM.

\noindent \textbf{Effect of Number of Related VMs.}
With the integration of extra inputs and multiple frequencies feature extraction, our T-ResNet models achieve the best RMSE, MAE, and MAPE across two datasets. Specifically, we attempt three variants of our model with setting different numbers of the relevant VMs. The results show that introducing extra inputs indeed help the model improve prediction accuracy, which confirms our observations. However, this effect is not continuously increasing. When the extra input dimension reaches the threshold, the earnings of improvement will decrease. The reason is that introducing more extraneous inputs will also introduce noise.

\section{Related Work} \label{sec: related}
\vspace{-.06in}
\noindent \textbf{VM Workload Prediction.}
With the development of cloud technology, many researchers have focused on predicting the workloads of VMs. Calheiros et al.~\cite{calheiros2015workload} paid attention to predicting the requests from web servers, but the data does not really reflect the consumption of workload. Farahnakian et al.~\cite{farahnakian2013lircup} used linear regression method to predict the CPU utilization based on past one-hour data. Such a method can only approximate short-term CPU utilization. Khan et al.~\cite{khan2012workload} discovered repeatable workload patterns within groups of VMs that belong to a cloud customer and further designed an HMM-based method to predict the changes of these patterns. However, the method has restricted the states of workload levels and cannot model new states. Some works~\cite{islam2012empirical, xue2015practise} focused on using neural network based model to predict the VM workload. Islam et al.~\cite{islam2012empirical} employed the extra sliding window technique
to evaluate the impact of different windows on the prediction accuracy. Xue et al.~\cite{xue2015practise} trained a group of networks models and further generate the prediction based on ensemble of these pre-trained networks. 
These works are different from ours where the proposed methods cannot consider both characteristics of VMs and effective algorithms that can capture specified temporal relations. 

\noindent \textbf{Time Series Prediction.}
Time series forecasting is the most common problem and we have many general methods to handle it.
ARIMA is popular and widely used statistical method for time series forecasting, and it combines autoregression and differencing together to model linear relationship~\cite{makridakis1997arma}. Hidden Markov model describes the process that 
the observations are produced by corresponding hidden states which randomly generated by hidden Markov chain before~\cite{khan2012workload}. 
The Markov process is based on finite pre-defined states and can capture limited nonlinear patterns. Recurrent neural network (RNN)~\cite{elman1990finding} is well designed neural network for sequence tasks, but vanilla RNNs suffer from gradient vanish problem~\cite{bengio1994learning}. Long short-term memory units (LSTM)~\cite{hochreiter1997long} and gated recurrent units (GRU)~\cite{cho2014learning} are proposed to handle this problem and keep long-term dependencies by memory cell. Moreover, convolutional neural network (CNN) has exhibited strong ability in image recognition which benefits from recent AlexNet~\cite{krizhevsky2012imagenet}. Prior works~\cite{van2016wavenet,cui2016multi} also showed the effectiveness of one-dimensional convolution in solving sequence problems. Neural networks with stacked convolutions can learn more complex patterns, and residual networks~\cite{he2016deep} make the learning of deep models possible. While RNN and CNN have the ability to capture nonlinear relations, our proposed model and experiments validate the view that networks are still hard to learn rules without specified structures.

\section{Conclusion} \label{sec:conclusion}
\vspace{-.06in}
In this paper, we studied the problem of VM CPU utilization prediction, which is an important task for cloud resource managers. We conduct an empirical analysis over real-world VM CPU utilization data, showing that deployment-based VMs have internal and external CPU utilization features. The internal features imply that the CPU utilization series has periodicity, tendency, and locality, while the external features reflect that VMs in a deployment tend to work in parallel and their CPU utilization behaviors are similar. Based on the observations, we propose Time-aware Residual Networks model named T-ResNet for prediction. 
T-ResNet consists of three residual networks to capture features at different frequencies and uses fully-connected layers to model deep feature interactions.
The experimental results verify the effectiveness of our proposed method in terms of prediction accuracy, compared with various baseline approaches.

\noindent {\bf Acknowledgements.} This research is supported in part by 973 Program (no. 2014CB340303), NSFC (No. 61772341, 61472254, 61572324, 61170238, 61602297 and 61472241) and the Shanghai Municipal Commission of Economy and Informatization
(No. 201701052). This work was also supported by the Program for Changjiang Young Scholars in University of China, the Program for China Top Young Talents, and the Program for Shanghai Top Young Talents.

\bibliographystyle{splncs}
\bibliography{myref}

\begin{thebibliography}{10}

\bibitem{gusev2013cpu}
Gusev, M., Ristov, S., Simjanoska, M., Velkoski, G.:
\newblock Cpu utilization while scaling resources in the cloud.
\newblock Cloud Computing (2013)

\bibitem{herodotou2011starfish}
Herodotou, H., Lim, H., Luo, G., Borisov, N., Dong, L., Cetin, F.B., Babu, S.:
\newblock Starfish: a self-tuning system for big data analytics.
\newblock In: Cidr. (2011)

\bibitem{clark2005live}
Clark, C., Fraser, K., Hand, S., Hansen, J.G., Jul, E., Limpach, C., Pratt, I.,
  Warfield, A.:
\newblock Live migration of virtual machines.
\newblock In: NSDI. (2005)

\bibitem{calheiros2015workload}
Calheiros, R.N., Masoumi, E., Ranjan, R., Buyya, R.:
\newblock Workload prediction using arima model and its impact on cloud
  applications’ qos.
\newblock TCC \textbf{3}(4) (2015)

\bibitem{farahnakian2013lircup}
Farahnakian, F., Liljeberg, P., Plosila, J.:
\newblock Lircup: Linear regression based cpu usage prediction algorithm for
  live migration of virtual machines in data centers.
\newblock In: SEAA. (2013)

\bibitem{gong2010press}
Gong, Z., Gu, X., Wilkes, J.:
\newblock Press: Predictive elastic resource scaling for cloud systems.
\newblock In: CNSM. (2010)

\bibitem{khan2012workload}
Khan, A., Yan, X., Tao, S., Anerousis, N.:
\newblock Workload characterization and prediction in the cloud: A multiple
  time series approach.
\newblock In: NOMS. (2012)

\bibitem{islam2012empirical}
Islam, S., Keung, J., Lee, K., Liu, A.:
\newblock Empirical prediction models for adaptive resource provisioning in the
  cloud.
\newblock FGCS \textbf{28}(1) (2012)

\bibitem{cortez2017resource}
Cortez, E., Bonde, A., Muzio, A., Russinovich, M., Fontoura, M., Bianchini, R.:
\newblock Resource central: Understanding and predicting workloads for improved
  resource management in large cloud platforms.
\newblock In: SOSP. (2017)

\bibitem{he2016deep}
He, K., Zhang, X., Ren, S., Sun, J.:
\newblock Deep residual learning for image recognition.
\newblock In: CVPR. (2016)

\bibitem{cleveland1990stl}
Cleveland, R.B., Cleveland, W.S., Terpenning, I.:
\newblock Stl: A seasonal-trend decomposition procedure based on loess.
\newblock Journal of Official Statistics \textbf{6}(1) (1990)

\bibitem{silverman2018density}
Silverman, B.W.:
\newblock Density estimation for statistics and data analysis.
\newblock Routledge (2018)

\bibitem{zhang2017deep}
Zhang, J., Zheng, Y., Qi, D.:
\newblock Deep spatio-temporal residual networks for citywide crowd flows
  prediction.
\newblock In: AAAI. (2017)

\bibitem{rumelhart1985learning}
Rumelhart, D.E., Hinton, G.E., Williams, R.J.:
\newblock Learning internal representations by error propagation.
\newblock Technical report, California Univ San Diego La Jolla Inst for
  Cognitive Science (1985)

\bibitem{lecun1998gradient}
LeCun, Y., Bottou, L., Bengio, Y., Haffner, P.:
\newblock Gradient-based learning applied to document recognition.
\newblock Proceedings of the IEEE \textbf{86}(11) (1998)

\bibitem{van2016wavenet}
Van Den~Oord, A., Dieleman, S., Zen, H., Simonyan, K., Vinyals, O., Graves, A.,
  Kalchbrenner, N., Senior, A., Kavukcuoglu, K.:
\newblock Wavenet: A generative model for raw audio.
\newblock arXiv preprint arXiv:1609.03499 (2016)

\bibitem{cui2016multi}
Cui, Z., Chen, W., Chen, Y.:
\newblock Multi-scale convolutional neural networks for time series
  classification.
\newblock arXiv preprint arXiv:1603.06995 (2016)

\bibitem{glorot2010understanding}
Glorot, X., Bengio, Y.:
\newblock Understanding the difficulty of training deep feedforward neural
  networks.
\newblock In: AIStats. (2010)

\bibitem{he2016identity}
He, K., Zhang, X., Ren, S., Sun, J.:
\newblock Identity mappings in deep residual networks.
\newblock In: ECCV. (2016)

\bibitem{ioffe2015batch}
Ioffe, S., Szegedy, C.:
\newblock Batch normalization: Accelerating deep network training by reducing
  internal covariate shift.
\newblock In: ICML. (2015)

\bibitem{nair2010rectified}
Nair, V., Hinton, G.E.:
\newblock Rectified linear units improve restricted boltzmann machines.
\newblock In: ICML. (2010)

\bibitem{kingma2014adam}
Kingma, D.P., Ba, J.:
\newblock Adam: A method for stochastic optimization.
\newblock arXiv preprint arXiv:1412.6980 (2014)

\bibitem{polyak1964some}
Polyak, B.T.:
\newblock Some methods of speeding up the convergence of iteration methods.
\newblock USSR Computational Mathematics and Mathematical Physics \textbf{4}(5)
  (1964)

\bibitem{Tieleman2012}
Tieleman, T., Hinton, G.:
\newblock {Lecture 6.5---RmsProp: Divide the gradient by a running average of
  its recent magnitude}.
\newblock COURSERA: Neural Networks for Machine Learning (2012)

\bibitem{makridakis1997arma}
Makridakis, S., Hibon, M.:
\newblock Arma models and the box--jenkins methodology.
\newblock Journal of Forecasting (1997)

\bibitem{chen2016xgboost}
Chen, T., Guestrin, C.:
\newblock Xgboost: A scalable tree boosting system.
\newblock In: SIGKDD. (2016)

\bibitem{hochreiter1997long}
Hochreiter, S., Schmidhuber, J.:
\newblock Long short-term memory.
\newblock Neural computation \textbf{9}(8) (1997)

\bibitem{xue2015practise}
Xue, J., Yan, F., Birke, R., Chen, L.Y., Scherer, T., Smirni, E.:
\newblock Practise: Robust prediction of data center time series.
\newblock In: CNSM. (2015)

\bibitem{elman1990finding}
Elman, J.L.:
\newblock Finding structure in time.
\newblock Cognitive science \textbf{14}(2) (1990)

\bibitem{bengio1994learning}
Bengio, Y., Simard, P., Frasconi, P.:
\newblock Learning long-term dependencies with gradient descent is difficult.
\newblock IEEE transactions on neural networks \textbf{5}(2) (1994)

\bibitem{cho2014learning}
Cho, K., Van~Merri{\"e}nboer, B., Gulcehre, C., Bahdanau, D., Bougares, F.,
  Schwenk, H., Bengio, Y.:
\newblock Learning phrase representations using rnn encoder-decoder for
  statistical machine translation.
\newblock arXiv preprint arXiv:1406.1078 (2014)

\bibitem{krizhevsky2012imagenet}
Krizhevsky, A., Sutskever, I., Hinton, G.E.:
\newblock Imagenet classification with deep convolutional neural networks.
\newblock In: NIPS. (2012)

\end{thebibliography}

\end{document}